\documentclass[ancac3]{achemso}
\setkeys{acs}{usetitle = true}
\usepackage{graphicx}
\usepackage{multirow}
\usepackage{gensymb}
\usepackage{epsfig}

\title{Chiral `Pinwheel' Heterojunctions Self-Assembled from C$_{60}$ and Pentacene}
\author{Joseph A. Smerdon}
\affiliation{Department of Physics, University of Liverpool, L69 3BX, UK}
\author{Rees B. Rankin}
\author{Jeffrey P. Greeley}
\author{Nathan P. Guisinger}
\author{Jeffrey R. Guest}
\email{jrguest@anl.gov}
\affiliation{Center for Nanoscale Materials, Argonne National Laboratory, Argonne IL 60439, USA}

\begin{document}
\begin{abstract}
\singlespacing
We demonstrate the self-assembly of C$_{60}$ and pentacene (Pn) molecules into acceptor-donor heterostructures which are well-ordered and -- despite the high degree of symmetry of the constituent molecules --  {\it chiral}.  Pn was deposited on Cu(111) to monolayer coverage, producing the random-tiling ($R$) phase as previously described.  Atop $R$-phase Pn, post-deposited C$_{60}$ molecules cause rearrangement of the Pn molecules into domains based on chiral supramolecular `pinwheels'.  These two molecules are the highest-symmetry achiral molecules so far observed to coalesce into chiral heterostructures.  Also, the chiral pinwheels (composed of 1 C$_{60}$ and 6 Pn each) may share Pn molecules in different ways to produce structures with different lattice parameters and degree of chirality.  High-resolution scanning tunneling microscopy (STM) results and knowledge of adsorption sites allow the determination of these structures to a high degree of confidence.  The measurement of chiral angles identical to those predicted is a further demonstration of the accuracy of the models. Van der Waals density functional theory calculations reveal that the Pn molecules around each C$_{60}$ are torsionally flexed around their long molecular axes and that there is charge transfer from C$_{60}$ to Pn in each pinwheel.

\end{abstract}

\maketitle
\newpage

Chiral surfaces and chiral surface structures have received a great deal of attention in recent years \cite{Barlow03-ssr,Ernst06-tcc}, both as powerful platforms for investigating the importance of symmetries in self-organization and chemical reactivity and for the potential technological dividends that they may pay.  Life on Earth is enantiospecific; not only is the origin of this handedness a persistent mystery, the chiral nature of many biomolecules is the reason that the physiological reaction to different enantiomeric forms of the same molecule can range from beneficial to inactive or even toxic.  This highlights the importance of developing systems which can perform enantio-selective catalysis or differentiate and/or separate enantiomers from one another.  In addition, chiral molecules or structures can manifest their handedness in optical interactions, leading to circular dichroism, polarization-sensitive photoemission~\cite{Polcik04-prl}, and allowed second-order nonlinear optical response~\cite{Verbiest98-science, Mulligan05-acie}.  Two dimensional systems, \textit{i.e.} surfaces, offer an important route to develop and explore these systems; structural freedom dwindles from 230 space groups in 3D to 5 chiral groups in 2D \cite{Raval02-jpcm}.   The introduction of chirality into the structure of self-assembled molecular systems provides a means of introducing a handedness to their functioning.

Chirality at surfaces is common, as intrinsically chiral materials such as quartz have chiral surfaces.  In addition to these examples, any atomic epitaxial system in which the overlayer breaks the symmetry of the substrate is chiral, such as the commonly observed $(\sqrt{5}\times\sqrt{5})R26.7^\circ$ superstructure on an fcc(100) surface \cite{Chen01-ss} and $(\sqrt{7}\times\sqrt{7})R19^\circ$ superstructure on an fcc(111) surface \cite{Hubbard83-jecie}.  The pattern of kinks in the step edges of high-index crystal surfaces \cite{Baber08-jpcc} can be used to seed chiral surfaces, but these systems are often difficult to prepare due to the general degree of disorder of such surfaces borne from the low coordination of surface atoms compared to alternative low-index terminations.  A much easier solution is to use molecular adsorbates to `nanopattern' a simple surface \cite{Pawin06-s,Theobald03-nature,Corso04-science,Barth05-nature,Wan06-acr,Stepanow06-cc,Zhang07-small,Li08-acr,Zhang08-jacs,Chen08-apl} to add chiral selectivity \cite{Lorenzo00-nature,Humblot02-jacs,Zhao00-jacs}.  It is possible to use chiral molecules to form chiral surfaces; a racemic mixture of enantiomers may undergo separation upon adsorption to form enantiopure domains.  \emph{Prochiral} molecules, which are achiral in the gas phase, develop surface-confined chirality upon adsorption on a 2D surface~\cite{Sowerby96-jme,Yablon00-jpcb,Humblot05-ass,Schock06-jpcb}.  Homochiral 2D chirality can involve unit cells from one up to many molecules including ordering in which chiral arrangements of single molecules hierarchically self assemble to produce large-scale ordered chiral structures \cite{Ecija10-acsn}. When adsorbed on Bi(111), arrays of around 45 achiral pentacene (Pn) molecules form homomolecular chiral pinwheels comprising six vanes of $\pi$-stacked sideways-laying molecules\cite{Sun12-small}.
Polymers composed of achiral modules were also recently demonstrated to be capable of adopting chiral structures based on the flexibility of the molecules combined with the module conformation with the surface \cite{Heim10-jacs}.

Pentacene and carbon-60 (C$_{60}$) are two archetypal molecules for the development of organic electronics and opto-electronics, boasting extremely high electron and hole mobilities, strong light absorption, and electron affinities which make them well-suited as donor and acceptor components of organic heterojunctions \cite{Horowitz91-sm,Haddon95-apl}.  When deposited on low-index metal faces, they are also archetypal systems for the study of self-assembly and generally form long-range achiral ordered networks.  This is in large part due to their high degree of symmetry; C$_{60}$ is a spherical achiral molecule with $I_h$ (icosahedral) symmetry, and Pn is a planar achiral aromatic molecule comprising five acene rings linearly fused along C-C bonds with $D_2$ symmetry.

When adsorbed to single-layer coverage on single-crystal metal surfaces such as Ag(111), Au(111) and Cu(111), Pn forms well-ordered monolayers with the molecular plane parallel to the surface \cite{Dougherty08-jpcc,Eremtchenko05-prb,Kafer07-cpl,Kang03-apl,France02-nl,Smerdon11-prb}.  Heterostructures involving Pn and C$_{60}$ on Ag(111) have been reported, with the C$_{60}$ molecules forming `nanochains' between rows of Pn molecules in a Pn bilayer \cite{Dougherty08-prb}. Also, chiral heterostructures including Pn on Ag(111) have been reported \cite{Chen08-jacs}.  It has also been recently shown that in-plane heterolayers with highly tunable structural parameters may be formed by varying the proportions of molecules sequentially deposited on an inert HOPG surface \cite{Huang10-small}.  Often, segregated domains of acceptor and donor molecules are formed following codeposition, as is observed for ex-TTF (donor) and PCBM (acceptor) molecules on Au(111) \cite{Otero07-nl}.  The use of precursor molecular layers to pattern co-adsorbed C$_{60}$ is described in various reports \cite{Yoshimoto07-jpp,Xu10-apl,Saywell08-jpcc,Wang11-jap}.  It has been demonstrated that when a molecule with a lower degree of symmetry than Pn (acridine-9-carboxylic acid (ACA)) is adsorbed on Ag(111) prior to C$_{60}$, domains of chiral heterostructures can be formed \cite{Xu05-nl}. 

In this paper, we demonstrate the self-assembly of these highly symmetric and achiral acceptor and donor molecules into two different types of {\it chiral} heterojunctions, in which the degree of chirality and the lattice parameters of the resulting structures may be varied.   To our knowledge, this is the most extreme example of symmetry-breaking observed on surfaces to date, where both molecules remain achiral if they are isolated on a surface -- remaining achiral to monolayer coverage -- but surrender this symmetry in these chiral structures.  We present structure solutions for these phases that fit the data extremely well.  We also present density functional theory (DFT) calculations that quantify the stability and electronic behavior of these heterolayers and precursors and confirm that these chiral networks are low-energy states of the system.  We also find that the constituent donor and acceptor molecules undergo charge transfer.

\section{Results and Discussion}
\subsection{STM Measurements and Morphology of Pinwheels}

In earlier work, we showed that Pn molecules can adopt various structures on Cu(111) depending on their molecular density on the surface~\cite{Smerdon11-prb}. The Pn molecules in these structures adsorb with the central benzene ring atop hcp hollow sites in the Cu(111) surface \cite{Lagoute04-prb} and oriented in a planar fashion along a Cu close-packed direction.  The two structures of particular interest here are the highly-ordered and dense herringbone ($H$) Pn (0.048 Pn molecules per surface Cu atom) and the lower-density random-tiling ($R$) Pn (0.034 Pn molecules per surface Cu atom).  C$_{60}$ adsorption on $H$-phase Pn results in a two-layered mostly disordered heterostructure (shown in Supplemental Information), with small domains of at most 10 fullerene molecules occupying sites consistent with the underlying Cu(111)--(6 3, 0 7)--2Pn structure.  

\begin{figure*}
\begin{center}
\includegraphics[width=1\textwidth]{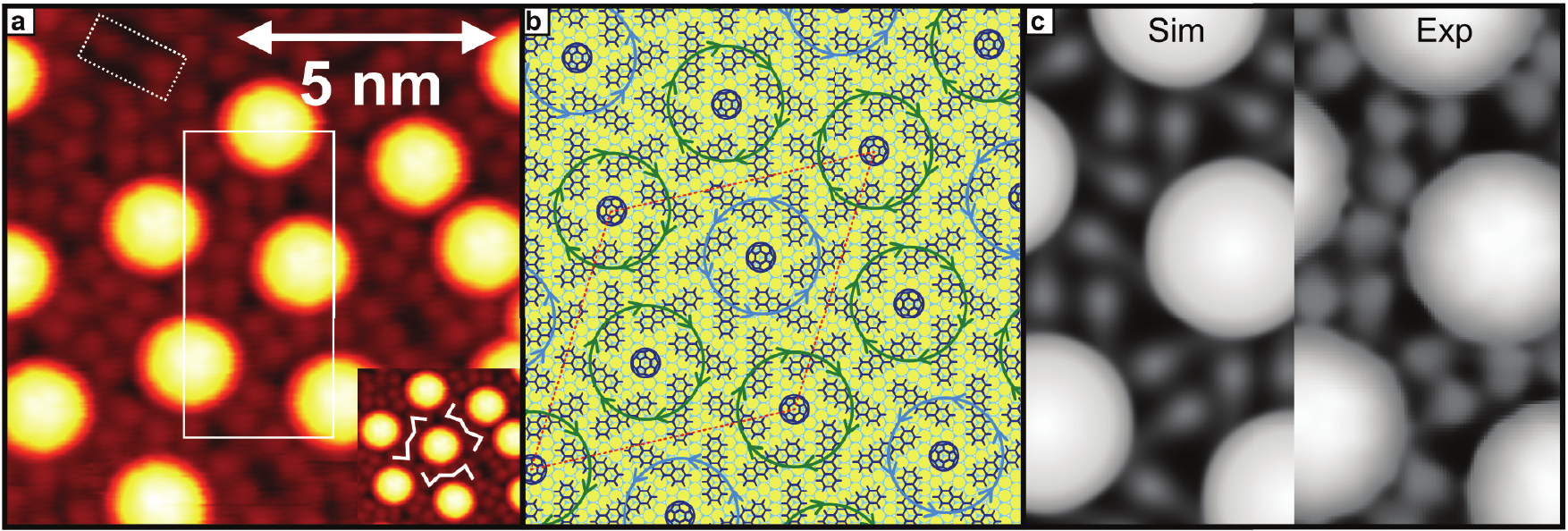}
\caption{\emph{(Color online)} \textbf{(a)}; 10 nm $\times$ 10 nm STM topograph ($V_{gap} = 2$ V, $I_T = 1$ nA) of C$_{60}$/Pn/Cu(111), showing a small hexagonal domain with a large lattice parameter.  An individual pentacene molecule is indicated with a dotted box. \emph{Inset}; a portion of the image with features connected by superimposed lines to aid identification of the chirality of the structure; \textbf{(b)}; a model of the C$_{60}$-Pn heterostructure based on a tiling of both chiralities of pinwheels of Pn molecules, with the unit cell of the structure indicated; \textbf{(c)}; \emph{Left}: The model in (b) plotted with representations of the molecules as they appear under STM; \emph{Right}: A detail of the experimental data, indicated in panel (a).}
\label{sparser}
\end{center}
\end{figure*}

When the bimolecular film was provided more space to reorganize by depositing C$_{60}$ atop the slightly lower-density $R$-phase Pn, a very different set of structures emerges.  An STM topograph of this film is reproduced in \ref{sparser} (a).  A hexagonal domain is clearly visible in which C$_{60}$ molecules are separated by $^\sim$3~nm and spaced by repeating patterns of Pn molecules.  The sparser $R$-phase Pn layer has been observed to permit tip-induced motion for molecules located in sites with adjacent vacancies \cite{Smerdon11-prb}.  The greater interaction strength between Cu and C$_{60}$ appears to drive a reorganization of the heterolayer resulting in the formation of these C$_{60}$-6Pn heptamer `pinwheels'.  While the C$_{60}$ molecules in \ref{sparser} (a) appear $^\sim$2~nm across and thus occlude much of the Pn between them, the raised ends of the visible Pn molecules are arranged in a chiral fashion as indicated in the inset of \ref{sparser} (a), suggesting that the structure itself is chiral. The images have been combined with previous observations of the various packings of Pn molecules \cite{Smerdon11-prb} on the Cu(111) surface to determine a unique structure, which is shown schematically with the underlying Cu(111) in \ref{sparser} (b).  We arrive at this structure by allowing each molecule to adopt a previously observed preferred nucleation site, which for Pn has the molecule oriented with the long axis along a close-packed Cu $[1\bar{1}0]$ direction and for both molecules has the molecule center above an hcp hollow site with a $b$-plane atom beneath \cite{Lagoute04-prb,Larsson08-prb}.  Each pinwheel consists of a central C$_{60}$ molecule surrounded by six Pn molecules.   This structure naturally results in reasonable separations for Pn molecules, with the overall Pn distribution separable into few-molecule domains of the $H$-phase structure in 3 orientations. In \ref{sparser} (c), we compare the experimental data with a version of the model plotted with representations of the molecules as they appear under STM, i.e. a large sphere for fullerene and a dumb-bell shape for pentacene.  This simple model shows a good agreement with the detail from the experimental data, plotted alongside.

Each individual pinwheel heptamer in \ref{sparser} (b) is \emph{chiral}; blue circles indicate right-handed (RH) pinwheels where the Pns are canted in the counter-clockwise direction, and green circles indicate left-handed (LH) pinwheels where the Pns are canted in the clockwise direction.  The unit cell of this superstructure consists of 1 right-handed (RH) pinwheel and 2 left-handed (LH) pinwheels, leading to an overall RH chirality in this kind of structure.  The matrix notation for the structure in \ref{sparser} (c) is Cu(111)--(16 -7, 23 16)--18Pn+3C$_{60}$.  The notation for the opposite chirality (a structure with twice the number of RH pinwheels as LH) is Cu(111)--(-7 16, 16 23)--18Pn+3C$_{60}$.  Based on the lattice parameter for Cu (0.3615 nm) and thus the nearest neighbor distance for Cu(111) (Cu$_{NN}=\frac{0.3615}{\sqrt{2}}=0.2556$ nm), expected lattice parameters for these structures are calculated and compared with measured lattice parameters in \ref{table}.  The density of Pn molecules in this structure is 0.043 per surface Cu atom, which is significantly higher than that observed in $R$-phase Pn despite the additional C$_{60}$ molecules and approaches the density of $H$-phase Pn, which is 0.047.  The total molecular density is 0.050 molecules per surface Cu atom.

\begin{figure}
\begin{center}
\includegraphics[width=0.5 \textwidth]{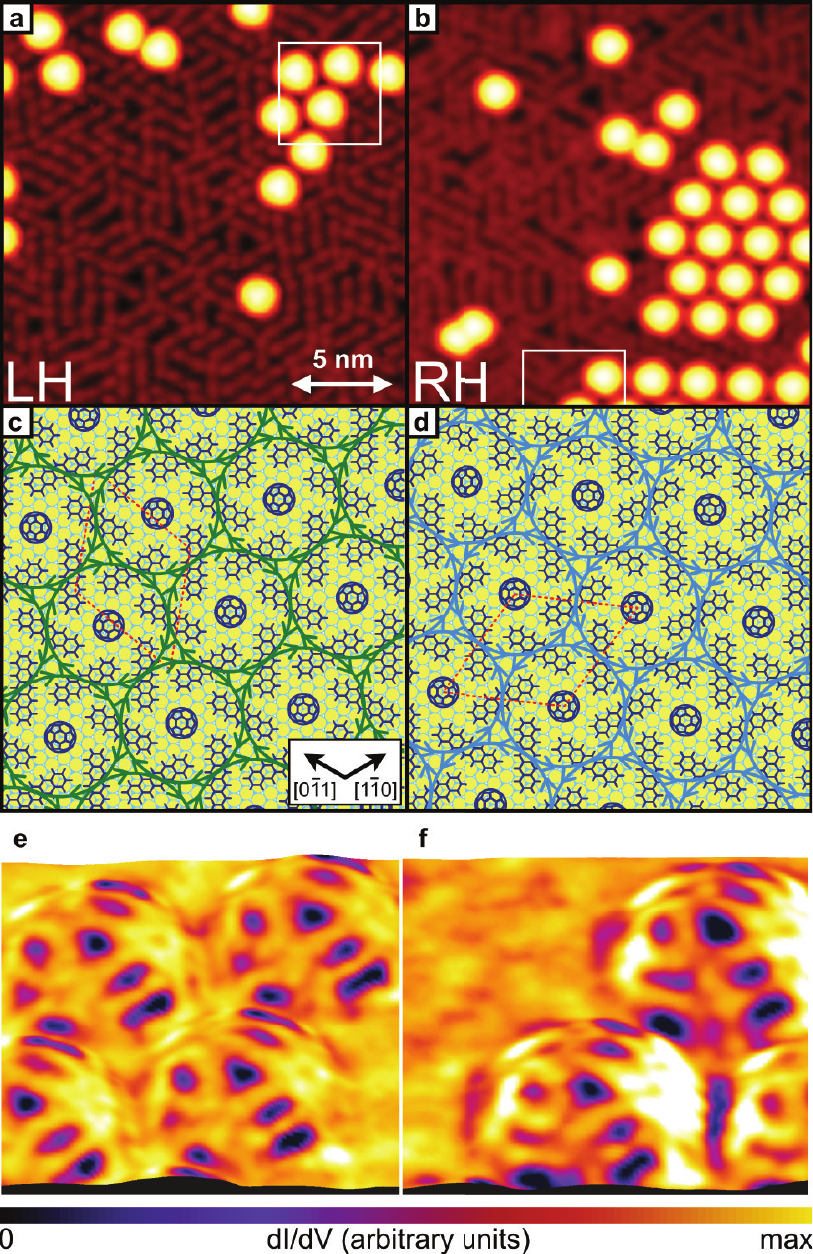}
\caption{\emph{(Color online)}\textbf{(a)}; 20 nm $\times$ 20 nm STM topograph ($V_{gap} = 1.2$ V, $I_T = 0.5$ nA) of C$_{60}$/Pn/Cu(111), showing an hexagonal domain; \textbf{(b)}; 20 nm $\times$ 20 nm STM topograph ($V_{gap} = 1.2$ V, $I_T = 0.5$ nA) of C$_{60}$/Pn/Cu(111), showing the opposite chirality to the domain portrayed in \emph{(a)}; \textbf{(c-d)}; models of the chiral C$_{60}$-Pn heterostructures  (c); \textbf{(e-f)}; 3D representations of portions indicated in \emph{(a-b)} of the topography textured with simultaneously acquired ($V_{gap} = 1.2$ V, $I_T = 0.5$ nA) dI/dV maps showing common C$_{60}$ orientation.}
\label{sqfigchi}
\end{center}
\end{figure}

\begin{figure}
\begin{center}
\includegraphics[width=0.5\textwidth]{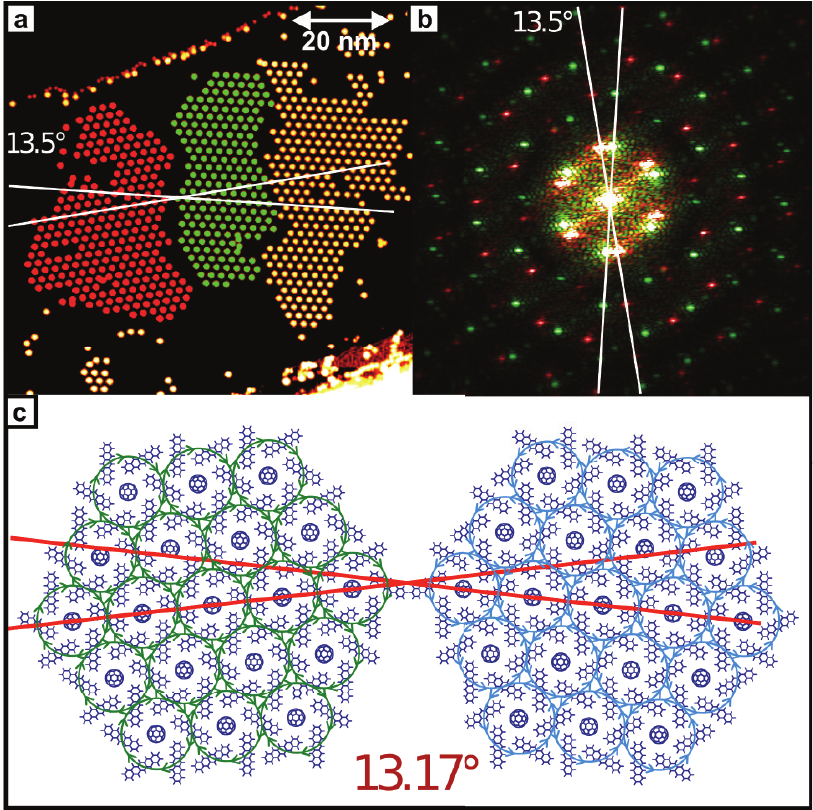}
\caption{\emph{(Color online)} \textbf{(a)}; 100 nm $\times$ 100 nm STM topograph ($V_{gap} = -1$ V, $I_T = 0.1$ nA) of C$_{60}$/Pn/Cu(111), showing some hexagonal domains of both chiralities. One domain is colored green and one red to correspond to the components of the Fourier transform in \textbf{(b)}; Fourier transforms of the domains indicated in \emph{(a)}, showing the angle between equivalent axes of the different domains to be $13.5^{\circ}$; \textbf{(c)}; models of the two opposite chiralities (Cu omitted for clarity), indicating the observed angle matches within experimental error the geometric value of $2\times\arctan(\frac{1}{5\sqrt{3}})=13.17^{\circ}$.}
\label{angles}
\end{center}
\end{figure}

If opposite ends of Pn molecules can be on separate pinwheels,  it should be possible to form domains with a smaller unit cell based on a pinwheels of a \emph{single} chirality.  Indeed, single chirality domains were formed and identified on the Cu(111) surface as shown by the images in \ref{sqfigchi}.  The two columns of the figure each depict domains comprising a single pinwheel chirality in which Pn molecules are shared between adjacent C$_{60}$ molecules; \ref{sqfigchi} (a) and (c) show the image and structure of the LH domains (unit cell Cu(111)--(6 -4, 4 10)--3Pn+C$_{60}$), where the Pns are canted in the clockwise direction, while \ref{sqfigchi} (b) and (d) show the image and structure of the RH domains (unit cell Cu(111)--(4 -6, 10 4)--3Pn+C$_{60}$), where the Pns are canted in the counter-clockwise direction.   Our measurements show that these single chirality domains may grow to a much larger size and contain 100s of chiral heterostructures; 50~nm domains were observed in our data, as shown in  \ref{angles}.  \ref{table} shows the predicted and measured lattice parameters and relative orientations of different chiral domains; the agreement is extremely good.  The predicted lattice parameters for these structures were calculated, as before, by reference to the Cu surface nearest neighbor distance.  The density of Pn molecules in this structure is 0.0415 per surface Cu atom, and the total molecular density is 0.055 per surface Cu atom.

The orientation of individual C$_{60}$ molecules at the centers of the pinwheels is determined by mapping their molecular orbitals through differential conductance measurements\cite{Lu03-prl,Larsson08-prb}, as shown in \ref{sqfigchi} (e) and (f).  These images were recorded with a bias voltage of 1.2~V (and tunneling current of 0.5~nA), and are thus sensitive to the LUMO+1 molecular orbital. The maps show the unique orientation of the C$_{60}$ molecules on the Cu(111) surface, by showing the pronounced three-fold character of the LUMO+1 orbital, where each lobe is associated with a pentagonal facet surrounding the upmost hexagonal facet \cite{Lu03-prl}.  We therefore determine that the pentagonal face of the C$_{60}$ is oriented along a close-packed direction on the fcc surface, independent of the pinwheel chirality.

The clear difference in the angle each uniquely chiral structure makes with high-symmetry directions on the Cu(111) surface allows the chirality to be easily identifiable in larger-scale topography.  \ref{angles} (a) shows three larger domains of C$_{60}$--6Pn pinwheels.  This image is almost completely drift free, as can be observed from the Fourier transforms shown in \ref{angles} (b).  This allows us to extract an accurate relative orientation of the two adjacent domains of opposite chirality, which can be seen to agree well with the proposed model structures in \ref{angles} (c).  The simplest way to extract the angles is by inspection of \ref{sqfigchi} (c) and (d); the periodicity of the pinwheels in the horizontal direction  {\bf i} is $5 \sqrt{3}$ Cu$_{NN}$ and in the vertical direction {\bf j} is 1~Cu$_{NN}$, which gives the Cartesian vector for the $a$ component of the unit cell of ${\bf a} = 5 \sqrt{3} {\bf i} -  {\bf j}$ for the RH domain, and ${\bf a} = 5 \sqrt{3} {\bf i} +  {\bf j}$ for the LH domain.  This leads to an angle difference between the domains of 13.17 degrees, as shown in \ref{angles}(c), and agrees well with the measured value in  \ref{angles} (a) of 13.5 degrees. This agreement between model and observation is striking given the complexity of the structure and their origins in very simple considerations: the known nucleation sites and orientations of the molecules on Cu(111) and the preferred configurations and spacing between Pn molecules on a Cu(111) surface.  The measurement of the angle between chiral domains can be considered a `checksum' of the veracity of the entire set of structure solutions.

\begin{table*}
\begin{center}
\begin{tabular}{lclclclclclcl}
\hline
\multirow{2}{*}{\textbf{System on Cu(111)}}     	& \multirow{2}{*}{\textbf{Chirality}} 			& \multicolumn{2}{c}{\textbf{Lattice parameter} (nm)} 								& \multicolumn{2}{c}{\emph{$a$} $\angle$ Cu$[1\bar{1}0]$}	\\
										&			& \textit{expected} 					& \textit{measured}					& \textit{expected} 					& \textit{measured} 					\\
\hline
\rule{0pt}{1.05em}
			
(6 -4, 4 10)--3Pn+C$_{60}$     		&  LH 		& 2.228  							& $2.15\pm0.05$ 					& $\arctan({\frac{2\sqrt{3}}{8}})=23.4^\circ$ 						& $24\pm2^\circ$						\\ 
(4 -6, 10 4)--3Pn+C$_{60}$     		& RH 		& 2.228 							& $2.15\pm0.05$ 					& $\arctan({\frac{2\sqrt{3}}{8}})=23.4^\circ$ 						& $24\pm2^\circ$						\\ 
(16 -7, 23 16)--18Pn+3C$_{60}$     	& LH+2RH & 5.219  							& $5.13\pm0.1$							& $\arctan({\frac{8\sqrt{3}}{15}})=42.7^\circ$ 						& $45\pm3^\circ$							\\
(7 -16, 16 23)--18Pn+3C$_{60}$     	& RH+2LH & 5.219 							& $5.13\pm0.1$  						& $\arctan({\frac{8\sqrt{3}}{15}})=42.7^\circ$ 						& $45\pm3^\circ$							\\
\hline
\end{tabular}
\end{center}
\caption{Properties of the chiral structures on Cu(111).  Measurements are made on STM topographs with low drift and are averages over multiple lattice parameters in all observed domain directions.  The angles measured for the chiral domains are derived from the relative angular orientation of chiral domains for the smaller structures and the orientation of domains relative to well-resolved Pn molecules for the larger structures. The difference in errors on lattice parameters and angles is based on the relative amount of data available.}
\label{table}
\end{table*}

The reorganization of the Pn and C$_{60}$ into pinwheels hinges on the freedom of defects to move through the Pn film, as shown previously by the tip-induced movement of Pn molecules \cite{Smerdon11-prb} and the lack of reorganization of the more densely-populated $H$-phase under C$_{60}$ adsorption.  The capacity of Pn molecules to rearrange themselves to form pinwheels around C$_{60}$ molecules and the resultant sparse hexagonal chiral domains of C$_{60}$--6Pn heptamers are reminiscent of the case for C$_{60}$/ACA/Ag(111) \cite{Xu05-nl,Xu06-jacs}.   For the C$_{60}$/ACA/Ag(111) system, the route to the formation of extended coherent hexagonal domains is the hydrogen bonding of adjacent ACA molecules, which provides a local `matching rule' that dictates the ultimate (chiral) film structure.  Although, due to the delocalization of a proton between the two O atoms on an ACA molecule, it is considered to have the symmetry of a water molecule and is thus achiral on a surface, the formation of hydrogen bonds must necessarily associate the proton with a particular O and in doing so impose chirality upon each individual molecule.  For the system under discussion, in which there is no opportunity for hydrogen bonding, it turns out that no such molecular interaction is needed and the analogous characteristic is the sharing of Pn molecules between adjacent pinwheels.  Once a C$_{60}$ molecule has induced the formation of the first chiral pinwheel, further growth is constrained by the Pn-sharing matching rule to produce an enantiopure domain.  Throughout this process, each individual molecule remains achiral and with an unchanged degree of symmetry.

\subsection{Density Functional Theory Calculations}

The electronic characteristic that defines a heterojunction is charge transfer between the donor and acceptor molecules. In order to quantify the degree of charge transfer, we have calculated the electronic charge density in the proposed geometric unit cell of Cu(111)--(6 -4, 4 10)--3Pn+C$_{60}$ and compared it to the electronic charge density of the constituents  alone; this charge density difference (CDD) reflects the interactions between the acceptor and donor molecules.  Below, we begin by examining the CDD \emph{without surface Cu atoms} to examine the symmetric structure and then compare these results with the charge transfer calculated for pinwheels adsorbed onto Cu in a relaxed morphology.  As we are primarily interested in electronic and structural behavior, we only examine a single chirality below (LH).

\begin{figure}
\begin{center}
\includegraphics[width=0.5\textwidth]{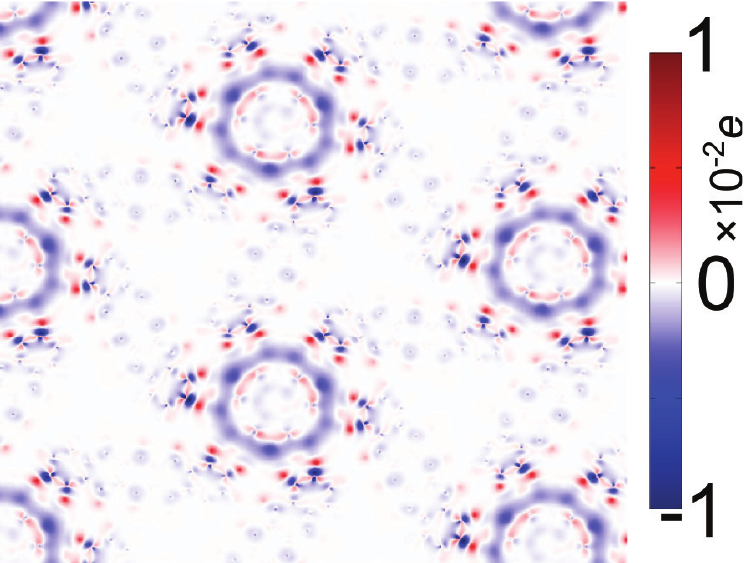}
\caption{\emph{(Color online)} The computed CDD for the unsupported (no Cu(111)) pinwheel structure at the average height of C atoms in the Pn network; positive CDD indicates positive charge accumulation relative to the reference constituent molecules.}
\label{rees1}
\end{center}
\end{figure}

For the unsupported model, the geometric structure of the pinwheel constituents was held fixed in its high-symmetry configuration.  The computed CDD of the assembled pinwheel compared to its constituents alone (in fixed geometry) is shown in \ref{rees1}. A slice plane of the CDD through the pinwheel is shown which is taken at the average height of C atoms in Pn molecules. Using Bader charge analysis, we find that each C$_{60}$ experiences a charge transfer of -0.08$e$ relative to the nominal valence.  The average charge transfer for a Pn molecule is +0.03$e$ relative to the nominal valence. From the results of both the CDD analysis and the Bader charge partitioning, it is clear that there is a small but existent charge transfer for the C$_{60}$-Pn pinwheel heterojunction in the absence of a Cu support which can be seen in the blue halo around the C$_{60}$ molecule in \ref{rees1}.

For our DFT calculations of the Cu-supported pinwheels, we seek to first duplicate some features of experimental observations of Pn monolayers on Cu(111). We begin with a low-coverage structure (shown in Supplemental Information figure S1) to establish the binding energy and charge transfer for isolated Pn/Cu(111). STM and AFM measurements~\cite{Smerdon11-prb,Lagoute04-prb,Gross09-science} have revealed some bowing of the Pn molecule on the Cu surface, where the ends of the molecule appear higher than the center.  While bowing of Pn/Cu(110) \cite{Muller09-prb} has been reproduced with DFT calculations, previous DFT calculations for Pn/Cu(111)~\cite{Toyoda09-jes} specifically excluded this from the calculation.  We observe a bowing of the molecule with the ends approximately 0.15 \AA~ higher than the central atoms in the Pn molecule.  We note that this bowing is less significant than observed on Pn/Cu(110) (0.4 \AA), and also that the bowing was only observed when VdW-corrected DFT calculations were employed.  We find a binding energy for the isolated Pn molecule on Cu(111) of 1.4 eV.  

Relaxation of the high density $H$-phase Pn structure results in bowing of all of the Pn molecules, as we previously observed experimentally \cite{Smerdon11-prb}.  The herringbone packing offers an additional $^{\sim}$0.2 eV stabilization to each Pn through attractive surface-mediated nearest-neighbor interactions.  Additionally, we have supplemented our electronic structure calculations with an analysis of charge transfer by using the Bader charge analysis tools as implemented in VASP by the Henkelman group\cite{Henkelman06-cms,Sanville07-jcc,Tang09-jpcm}. The charge transfer was determined to be $-0.10e$/Pn in the isolated Pn configuration and a slightly smaller $-0.08e$/Pn in the $H$-phase (interacting) configuration.

\begin{figure}
\begin{center}
\includegraphics[width=1\textwidth]{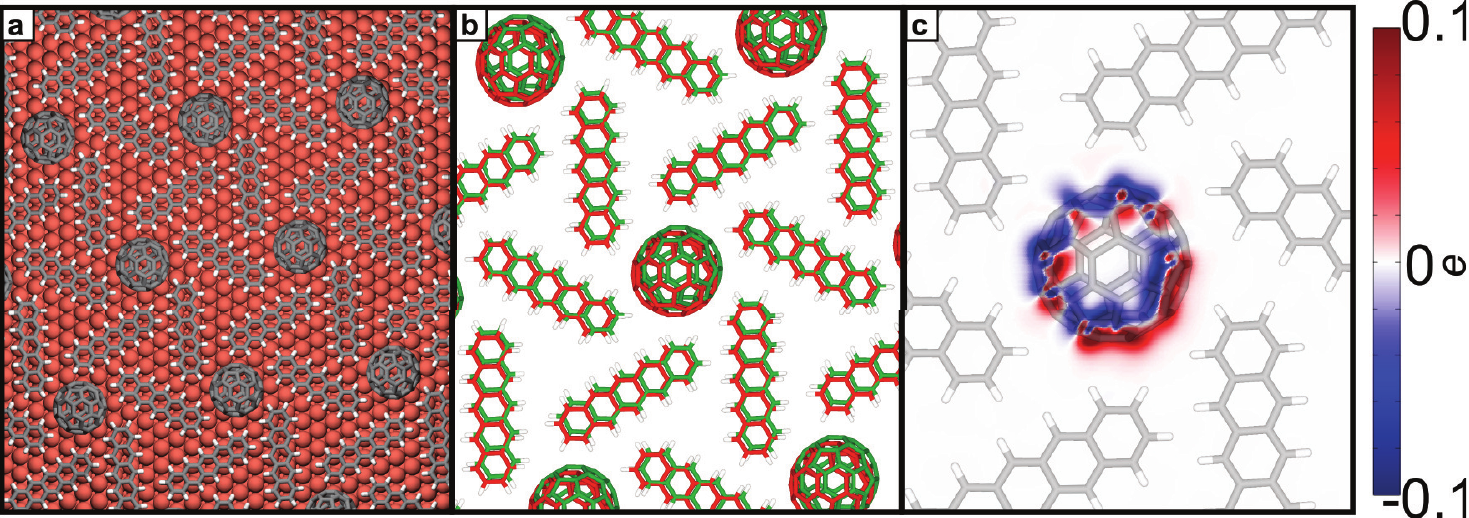}
\caption{\emph{(Color online)}   The relaxed geometry of the chiral pinwheel of 3Pn+C$_{60}$ on Cu(111) is shown \textbf{a)} in a top down view and \textbf{b)} with the Cu atoms removed (green) with a superimposed unsupported symmetric pinwheel discussed in the text and in  \ref{rees1} (red); \textbf{c)} the computed CDD for the relaxed chiral pinwheel on Cu(111) where the slice plane is taken at the same $z$-coordinate as in \ref{rees1}.  With the relaxation on and interaction with Cu(111), an electronic dipole forms in addition to the charge transfer.}
\label{rees3}
\end{center}
\end{figure}

We next perform calculations to characterize the properties of the Pn/C$_{60}$ pinwheels on Cu(111). In \ref{rees3}a, the computed low energy geometry of the pinwheel is shown in a top-down view on the Cu(111) surface itself. In the pinwheel, we observe some torsion of the Pn molecules around their long axes in addition to the bowing observed at the ends relative to the middle, as described for the isolated and $H$-phase Pn.  Also, the C$_{60}$ rotates and shifts slightly off its originally constructed high symmetry axes relative to the surface and close-packed rows. In \ref{rees3}b, a comparative structural diagram of the symmetric unsupported pinwheel geometry (red) is overlaid on the computed low-energy geometry of the pinwheel adlayer on Cu(111) (green) in a top-down view to highlight the shift from the high-symmetry site to one of the sixfold degenerate low-symmetry sites.

To understand the scale of the thermodynamic driving force for the formation of the chiral pinwheel network of Cu(111)--(6 -4, 4 10)--3Pn+C$_{60}$, we calculate the relative energetics of this structure and compare it to those of low-coverage isolated Pn molecules and to C$_{60}$ molecules on Cu(111). As mentioned previously, there is an appreciable interaction of approximately -0.2 eV per Pn to assemble $H$-phase Pn as compared to low coverage Pn on Cu(111).  Furthermore, there is a approximately +0.1 eV penalty per Pn to distort the $H$-phase structure into the arrangement that Pn occupies in the chiral pinwheels. Finally, the net thermodynamic gain to assemble chiral pinwheels of 3Pn+C$_{60}$ as compared to a stoichiometric equivalent of $H$-phase Pn and low coverage C$_{60}$ on Cu(111) is determined to be $^{\sim}$-0.35 eV per pinwheel.  Therefore, the calculations appear to agree with the observation that there is a thermodynamic driving force towards self-assembly of pinwheels at this stoichiometry rather than separate domains of C$_{60}$ and Pn.

We have (as before) carried out a Bader charge decomposition for the pinwheel structure which we compare to its constituent molecular species on Cu(111) - \ref{tabler} below.  The pinwheel structure can be seen to behave like a heterojunction; the acceptor C$_{60}$ becomes more negatively charged compared to the isolated C$_{60}$, while the Pn becomes more positively charged compared to the isolated Pn.  The calculations show that the Cu support helps enhance the charge transfer between Pn and C$_{60}$ as compared to the unsupported pinwheels, increasing its potential as a heterojunction.

\begin{table}[htdp]
\caption{Computed Bader charge ($e$) for Pn and C$_{60}$ on the Cu(111) surface in different structures. In this table, electron accumulation relative to nominal valence is denoted as negative. The value given is the average per Pn molecule.}
\begin{center}
\begin{tabular}{ccccc}
 & & {\bf Bader Charge ($e$)} & & \\
 \hline
Molecule & Pinwheel/unsupported & Isolated/Cu(111) & Pinwheel/Cu(111) & Herringbone/Cu(111) \\
\hline
C$_{60}$ & -0.08 & -0.09 & -0.18 & n/a \\
Pn & +0.03 & -0.10 & -0.03 & -0.08 \\
\hline
\end{tabular}
\end{center}
\label{tabler}
\end{table}

To further clarify the charge transfer in the dense chiral Pn/C$_{60}$ pinwheel on Cu(111), we have performed additional CDD analysis as shown in \ref{rees3}; the slice plane is taken at the average $z$-height of the C-atoms in the Pn molecules (same as in \ref{rees1}).  Beyond the scale of the CDD (enhanced by increased charge transfer due to the Cu(111) surface as discussed above), a fundamental difference is the fact that the computed CDD for the pinwheel adlayer structure on Cu(111) shows a net dipole rather than the six-fold symmetric CDD seen in \ref{rees1}. This is a consequence of the fact that the computed low-energy geometry acquires a minor asymmetry/tilt relative to the symmetric structure computed in the gas phase; indeed, the dipole is oriented along the tilt direction of the C$_{60}$.  As the Pn molecules in the pinwheel additionally move with the asymmetric placing of the C$_{60}$ molecule, this displacement may propagate throughout a pinwheel domain, which in turn may lock each domain into one of the 6 degenerate possibilities.  However, as the additional movement of the Pn molecules is very small, it is also possible that each C$_{60}$ molecule may behave nearly independently of the others.  We were unable to address this issue with our already computationally expensive single pinwheel basis set, and the effect is too subtle to prove or disprove with our experimental data. Also, if pinwheels can all shift independently, the thermal barrier to motion may be low enough that the pinwheel averages out \textit{via} rotation between the 6 possibilities.

Recent work has highlighted the switching behavior of C$_{60}$ molecules adsorbed on clean metallic surfaces and also on predeposited molecular networks \cite{Neel11-nl,Vijayaraghavan12-nl}. C$_{60}$ molecules switch between several stable orientations relative to the substrate when current is passed through them at a particular sample bias, leading to different molecular conductances.  Such switching behavior would seem to be compatible with the six degenerate possibilities for the C$_{60}$ positions, though as the molecular orientation is equivalent in each case it is unlikely that there would be an appreciable effect on overall conductance of the C$_{60}$ molecule.

\section{Conclusions}

In conclusion, we have studied the structures formed when Pn and C$_{60}$ are sequentially deposited on Cu(111).  When C$_{60}$ is deposited on $H$-phase Pn, a mostly disordered layer is observed with little rearrangement of the close-packed Pn layer beneath.  However, when C$_{60}$ is deposited on random-tiling Pn, rearrangement of the more loosely-packed layer proceeds with the formation of locally chiral domains.  Two different unit cell structures are identified, the smaller of which is observed in single chirality domains extending up to 50 nm in largest dimension for the coverages studied.  The variability of lattice parameters based on the relative arrangement of chiral pinwheels has not previously been observed, and it is possible that other structures may exist with larger lattice parameters and different populations of LH and RH pinwheels. 

Van der Waals corrected DFT calculations were performed to help quantify energetic, geometric and charge-transfer properties of the experimentally observed systems.  The calculations confirm there is a very appreciable net thermodynamic driving force to assemble the dense chiral pinwheel of 3Pn+C$_{60}$ on the Cu(111) surface. Further, the calculations show that the 3Pn/C$_{60}$ adlayer indeed shows charge transfer between Pn and C$_{60}$, and this effect is enhanced when the molecules are assembled in the pinwheel geometry on the Cu(111) surface.

As mentioned above, switching behavior has been observed for C$_{60}$ adsorption atop clean metallic surfaces and also atop a SAM.  It would be interesting to investigate how such switching behavior, if present, is modified by the additional interaction of the Pn molecules in this intermediate planar heterostructure system.

As Pn and C$_{60}$ are archetypal donor and acceptor molecules, the electronic properties of the chiral heterostructures which are observed are critically important and demand further investigation.  Whether these structures on Cu(111) are coupled strongly enough to provide a functional acceptor-donor heterojunction remains to be seen, but the self-assembly of these highly-symmetric molecules into chiral heterostructures suggests a template for realizing electronically active junctions with a handedness which should expand their potential technological impact.

\section{Methods}
\subsection{Experimental Details}
All measurements were carried out in a commercial Omicron UHV VT-AFM/STM operating with the sample maintained at 55 K.  The Cu(111) crystal was cleaned by simultaneously sputtering with Ar$^{+}$ ions at 1 keV and annealing to 900 K, with a final sputtering cycle at room temperature followed by an anneal at 900 K.  Pentacene and C$_{60}$ were deposited from a Dodecon 4-cell organic MBE source.  Pentacene was evaporated at 510 K, with a deposition rate on clean copper of approximately 0.2 ML per minute. Pentacene films were annealed post deposition as described below, therefore the deposition rate is approximate and does not necessarily bear a relation to the ultimate coverage.  Carbon-60 was evaporated at 730 K, with a deposition rate on a Pn film of approximately 0.07 ML per minute.  Deposition rates for each molecule were determined from subsequent STM topographs, by counting the number of molecules in a given area, comparing this to the number of molecules in a complete layer to find the coverage in ML. All gap voltages are quoted as sample bias.  Differential conductance (dI/dV) maps, collected with the standard lock-in technique ($f = 20$ kHz), are prepared by averaging forward scan and backward scan data to cancel the scan-induced artefacts from topography.  Tungsten tips were etched in 2 mol NaOH solution in a two stage process involving a coarse etch at 10 V followed by a fine etch under an optical microscope at 2 V.  When the tip picked up molecules during scanning -- which resulted in the appearance of multiple tips in the images -- it was $e$-beam heated at 2 W for 10 seconds; this process usually restored the imaging quality.

The $H$-phase ($R$-phase) was prepared by annealing a multilayer of Pn to 400 K (420 K) for 10 minutes.  Following this preparation and subsequent measurement of the Pn films at 55 K, the sample was warmed to ambient temperature and C$_{60}$ was deposited to approximately 0.1 ML coverage.

\subsection{Computational Details}

In order to facilitate understanding of the molecular-level structure of the chiral pinwheels on Cu(111), as well as their energetic stability and capability to serve as charge-transfer heterojunctions, we have employed plane-wave Density Functional Theory (DFT) calculations as implemented in the Vienna \emph{Ab-Initio} Simulation Package (VASP)\cite{Kresse93-prb,Kresse94-prb,Kresse96-prb,Kresse96-cms}.  The calculations used the PAW\cite{Blochl94-prb,Kresse99-prb} treatment of core electrons, together with the revPBE\cite{Perdew96-prl,Zhang98-prl} density functional.  Van der Waals corrections to the functional were implemented in the method of Langreth and Lundqvist\cite{Dion04-prl,Perez09-prl,Lee10-prb,Klimes11-prb}. The PREC=HIGH accuracy extension in VASP was employed with a cutoff energy of $^\sim$520 eV, FFT mesh-grids large enough to avoid wrap around errors, and a default Methfessel-Paxton\cite{Methfessel89-prb}  Fermi-smearing of 0.2 eV. The single gamma $k$-point was used in all calculations due to the very large system size(s). All calculations were performed fully spin-polarized, although it was not expected (or observed) that any species should acquire high spin states.  Generally good agreement between calculated and experimental lattice constants of bulk copper was observed: calculation (experiment) values were seen to be 3.67 \AA~(3.62 \AA).  A three layer thick (6 -4, 4 10) surface supercell was employed to study the high coverage 3Pn+C$_{60}$ pinwheel, and a (6 3, 0 7) surface supercell was studied for dense $H$-phase pentacene surface\cite{Smerdon11-prb} (for each case, the top layer on the adsorbate side was  allowed to relax). The vacuum spacing was set to approximately 20 \AA~ to ensure no $z$-axis interactions between neighboring cell images.  Geometric optimization of the near-surface atoms and adsorbates was performed with a force-convergence criteria of < 0.04 eV/\AA~ per atom, while self-consistent iteration on the electronic wavefunction was performed with a convergence criteria of at least $2.0\times10^{-7}$ eV.  For the adsorbates studied in this work, initial placement of molecules on the surface prior to geometric optimization was performed based on results \textit{via} location of the molecular centroid above the four high-symmetry sites (fcc, hcp, top, and bridge); due to the symmetry of both the adsorbates and the surface itself, no rotational dependence of the binding site was examined.  For isolated molecular species studied in the gas phase, a cubic supercell of 30\AA~$\times$30\AA~$\times$30\AA~ was used; for studies of the pinwheel network in the absence of a metal substrate (`gas phase'), the same unit cell was used as in the case of the adsorbed pinwheel network.  

\section{Acknowledgments}

This work was supported by the U.S. Department of Energy, Office of Science, Office of Basic Energy Sciences, under `SISGR' Contract No. DE-FG02-09ER16109.  Use of the Center for Nanoscale Materials was supported by the U. S. Department of Energy, Office of Science, Office of Basic Energy Sciences, under Contract No. DE-AC02-06CH11357.   J. Greeley acknowledges a DOE Early Career Award through the DOE Office of Science, Office of Basic Energy Sciences. The authors would like to acknowledge the technical assistance of B. Fisher, useful comments from E. Yitamben, L. Gao and N. Giebink and useful discussions with S.-W. Hla.

\section{Supporting Information}

Calculations for the relaxed herringbone pentacene atop Cu(111) structure and scanning tunneling microscopy data for an additional co-adsorption structure of pentacene and fullerene atop Cu(111).  This information is available free of charge \textit{via} the Internet at \url{http://pubs.acs.org}.

\providecommand*\mcitethebibliography{\thebibliography}
\csname @ifundefined\endcsname{endmcitethebibliography}
  {\let\endmcitethebibliography\endthebibliography}{}

\end{document}